# The Beam Halo Experiment at LEDA

P. L. Colestock, T. Wangler, C. K. Allen, R. L. Sheffield, D. Gilpatrick and the Diagnostics Group,
M. Thuot and the Controls Group, the LEDA Operations Team, Los Alamos National Laboratory,
M. Schulze and A. Harvey, General Atomics, Los Alamos, NM 87545

*Abstract*

Due to the potentially adverse effects of the generation of halo particles in intense proton beams, it is imperative to have a clear understanding of the mechanisms that can lead to halo formation for current and proposed high-intensity linacs. To this end a theoretical model has been developed, which indicates that protons under the combined influence of strong space charge forces and periodic focussing in a linear transport channel can be kicked into halo orbits. However, no experimental measurements of beam halo in proton beams have yet been carried out. In this paper we report the progress of an effort to carry out an experiment to measure beam-halo using the existing high-intensity proton beam of the LEDA facility. A linear transport channel has been assembled with the appropriate diagnostics for measuring the expected small beam component in the beam halo as a function of beam parameters. The experiment is based on the use of an array of high-dynamic-range wire and beam scrapers to determine the halo and core profiles along the transport channel. Details of the experimental design, the expected halo measurement properties will be presented.

## 1 INTRODUCTION

The interest in understanding the formation of a halo distribution around an intense proton beam has increased in recent years with the development of new applications requiring such beams. In order to understand this process, a theoretical model has been developed and extensive computer simulation has been carried out, reviewed elsewhere in these proceedings.[1] Although an extensive theoretical literature has evolved, there has been no definitive test of the model to date, owing in part to the fact that few intense proton beams of the required intensity exist.

At the Low Energy Demonstration Accelerator (LEDA)[2] at Los Alamos National Laboratory, we have embarked upon a program to carry out a first test of the halo formation model using the available intense proton beam from LEDA. A 52 quad transport line is being constructed following the RFQ of the LEDA injector, a 100 mA, 6.7 MeV proton beam with a capability for continuous operation. We will use, however, a pulsed beam with a 20 $\mu$sec pulse length and a $10^{-4}$ duty factor in order to facilitate the use of direct wire and scraper measurements of the beam profiles.

The purpose of this program is to make a detailed comparison, for the first time, between the theoretical model of halo formation and beam profiles in a controlled way.

## 2 EXPERIMENTAL SETUP

An overall view of the halo transport line is seen in Fig. 1. The transport channel consists of 52 quadrupoles with a G-l = 2.5 T, the first four of which have extra strength to permit mismatching the beam from the RFQ as it passes into the transport channel. An array of up to nine scanners will be used to monitor beam profiles in two dimensions over the length of the transport channel, and details of the scanner design are given elsewhere in these proceedings.[3] A scanner consists of 33 $\mu$ diameter carbon filaments that can safely intercept the entire beam current and can be passed through the core of the beam. Beam current is determined by secondary emission. On the same assembly a scraper plate is mounted which permits measurement of the diffuse halo region. Taken together, the beam profile measurements should permit a dynamic range of $10^5$.

The philosophy for scanner placement is the following: the first scanner is used to determine the beam distribution emerging from the RFQ, and this provides a critical initial condition on the halo evolution. The second group of four scanners is placed after quadrupole 13, which is where simulations show the beam is largely debunched. Since the transverse space charge forces depend on the longitudinal phase space density, it is reasoned that a fairly uniform domain for the halo formation should occur downstream from this point. Four scanners are used to ensure complete coverage of the phase space over a full betatron period (68 degrees per cell). A final array of four scanners at the end of the transport channel permits a similar measurement of phase space in the region where the halo is expected to be fully developed. Additional diagnostics include an array of beam position monitors, resistive-wall current monitors, current toroids and beam lossmonitors.

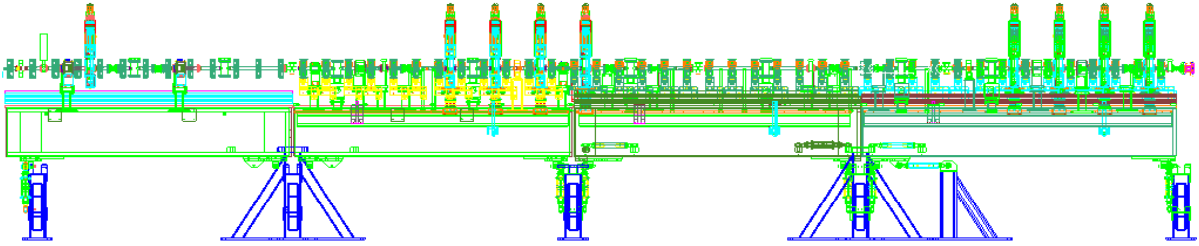

**Figure 1** Overall Layout of the Beam Halo Experiment on LEDA. The transport line consists of 52 quadrupoles between the LEDA RFQ and the HEBT/beam stop. The first four quadrupoles are used to create a controlled mismatch from the RFQ. A series of nine dual-axis, high dynamic-range wire/scraper scanners will be used to measure halo properties.

Details of the wire scanner design and implementation are given elsewhere in these proceedings[4], as well as aspects of the complex control system required for driving the scanners and acquiring scanner data.[5]

Beam operation will be limited to < 20 μsec macropulses to limit the power delivered to the scraper elements, and a software algorithm has been devised to prevent insertion of scrapers into the core of the beam. However, the combined use of wire filaments and scraper elements will permit a complete beam profile to be obtained. Profile data will rely on shot-to-shot repeatability of the RFQ, which has been measured to be about 1%. Position jitter was also measured to be a negligible effect.

Because of previously existing constraints on LEDA, the quadrupole magnets permit only a 3 cm bore beampipe, which is expected to be approximately 50% larger than the largest halo orbit. Moreover, the core beam size is about 1 mm RMS, which requires 5 mil alignment tolerances. This was carried out using a taut-wire system for magnet center fiducialization, and a precision alignment rail spanning the length of the transport line. Each magnet was individually mapped and aligned relative to the rail, along with beam position monitors. Load stresses and thermal expansion were measured to be negligible alignment factors. 1 – 2 mil tolerances were achieved. An array of steering magnets will be employed to correct for misalignment errors in conjunction with the eight beam position monitors. The magnets are powered in strings of eight, each with individual shunts for trim control. Four singlet supplies are used to individually power the match/mismatch magnets at the exit of the RFQ. Once the beam has exited the transport channel, it enters the high-energy beam transport line (HEBT) and terminates in the beam stop, both commissioned in a previous LEDA run.

## 3 EXPERIMENTAL OBJECTIVES

The experimental objective is to verify the halo formation model which requires a detailed comparison with simulations. This can, in principle, be obtained by the array of nine scanners and the data will be fitted to the results of beam simulations. Such an endeavor requires a massive data acquisition and simulation effort and is expected to be completed some months after data-taking has come to a conclusion. However, as noted in reference 2, a workable single-parameter measure of the halo formation is described based on the fourth moment of the distribution. This will serve as a rough measure of the existence of a halo. However, a complete confirmation of the model will require as complete a phase space picture as possible. Another signature of halo formation according to the theoretical model is the maximum extent of the halo particles. This value is achieved shortly after the mismatch quadrupoles, with only the number of particles within the halo distribution increasing along the transport channel. A measurement of this maximum particle radius using the sensitive scraper diagnostic, in particular as a function of the mismatch strength, will provide an important test of the theory.

An alternative approach is to vary the tune of the transport line effectively rotating the distribution past a set of scanners with fixed orientation. Since simulations show that both the core and halo distributions can be made to rotate rigidly over some phase range, it may be possible to use the an inverse Radon transform to form a detailed picture of phase space. Whether this elegant technique can be made to work in the experimental environment of an intense beam remains to be seen.

## 4 RUN PLAN

At the present time, all of the magnets and beamline have been installed, as well as most of the diagnostics hardware. Final checkout of the associated electronics and control systems has begun. Moreover, the LEDA RFQ and injector are being readied for operation.

Current run plans indicate that beam operation will commence shortly following this conference, with a commissioning and shakedown period for new hardware. Full data-taking will begin early in FY 01 and the experiment is expected to continue until Spring 2001.

Because of the necessity to handle large amounts of beam profile data, the initial emphasis will be placed on fully integrating the new diagnostic hardware into the EPICS-based control system.

Due to the potential for beam-induced damage at the full current available from the RFQ (100 mA), it is planned to implement a fast-protect system based on beam losses that will prevent any such damage. An extended period of operation at low currents (10 mA) is planned until the fast-protect system is in place, and all diagnostics have been checked out at low power densities.

## 5 SUMMARY

We have described the elements of an experiment to test the theoretical model for halo formation that is soon to be carried out on the LEDA facility. The purpose of the experiment is to verify the model with a detailed measurement of phase space along a linear transport channel, and by comparing this experimental information to analytical models and simulations. The combination of large dynamic-range profile diagnostics and large-scale simulations should provide a unique and comprehensive test of the theory.

## ACKNOWLEDGMENTS

The authors acknowledge support from the U.S. Department of Energy, and wish to give a special thanks to the many dedicated staff who have braved a fire and other challenges to put together this experiment in a short time.